\documentclass[graybox,natbib,nosecnum,11pt]{svmult}
\bibpunct{(}{)}{;}{a}{}{,} 

\pdfoutput=1   

\usepackage{mathptmx}       
\usepackage{type1cm}        
                            
\usepackage{tgschola}
\usepackage{hyperref}
\hypersetup{colorlinks,breaklinks,
            urlcolor=[rgb]{0,0,1.0},
            linkcolor=[rgb]{0,0,1.0}, 
            citecolor=[rgb]{0,0,1.0}}

\usepackage{makeidx}         
\usepackage{graphicx}        
\usepackage{multicol}        
\usepackage[bottom]{footmisc}
\usepackage[normalem]{ulem}	

\usepackage{soul}   

\newcommand*\aap{A\&A}

\newcommand*\apj{ApJ}
\newcommand*\apjl{ApJ}

\newcommand*\mnras{MNRAS}

\newcommand*\nat{Nature}

\newcommand*\pasp{PASP}

\newcommand{\omc}{($O-C$)}
\newcommand{\kms}{km s$^{-1}$}
\newcommand{\msun}{$M_{\odot}$}

\makeindex             

\begin{document}

\title*{Timing by Stellar Pulsations as an Exoplanet Discovery Method}
\author{J. J. Hermes}
\institute{Hubble Fellow, Department of Physics and Astronomy, University of North Carolina, Chapel Hill, NC\,-\,27599-3255, USA, \email{jjhermes@unc.edu}}
%
%
\maketitle

\abstract{The stable oscillations of pulsating stars can serve as accurate timepieces, which may be monitored for the influence of exoplanets. An external companion gravitationally tugs the host star, causing periodic changes in pulsation arrival times. This method is most sensitive to detecting substellar companions around the hottest pulsating stars, especially compact remnants like white dwarfs and hot subdwarfs, as well as $\delta$ Scuti variables (A stars). However, it is applicable to any pulsating star with sufficiently stable oscillations. Care must be taken to ensure that the changes in pulsation arrival times are not caused by intrinsic stellar variability; an external, light-travel-time effect from an exoplanet identically affects all pulsation modes. With more long-baseline photometric campaigns coming online, this method is yielding new detections of substellar companions.}

\section{Introduction}

Our sun is never fully at rest: it is constantly pirouetting around the center of mass of our solar system, tugged primarily by Jupiter and Saturn. Over the course of a few decades, the sun dances in a complicated orbit around the solar system barycenter, covering an effective diameter of more than 2 million km --- a distance it takes light more than 8 s to travel.

Timing techniques exploit the changing light-travel time of a variable star that is moved around a center of mass by an external companion, which can be an exoplanet or another star. Millisecond pulsars are some of the most precise clocks in the universe, and can be monitored for subtle shifts caused by orbiting bodies less massive than the moon (see Pulsar Timing as an Exoplanet Discovery Method). In a similar way, stellar pulsations can be observed for any periodic changes in their arrival times due to exoplanets.

An advanced alien civilization monitoring our sun for many years could search for the $8$ s light-travel-time effect imparted by the largest Jovian planets. (They would run into a problem: although the sun does in fact pulsate, the oscillations are not coherently driven, so they cannot be timed precisely enough to detect any of the planets in our solar system.) However, many other classes of pulsating stars have extremely regular oscillation periods that can be monitored with enough precision to detect the influence of substellar companions, especially A stars and compact remnants such as white dwarfs and hot subdwarfs.

In many cases, pulsation timing affords the best way to probe these stars for planets. For example, main-sequence A stars are difficult to probe with sufficient sensitivity using radial-velocity and transit techniques. However, a candidate exoplanet has been discovered around a $\delta$ Scuti star observed in the {\em Kepler} mission from the timing of its pulsations; this planet would exist near the habitable zone of its host A star \citep{2016ApJ...827L..17M}.

In this section we outline the technique of finding exoplanets using stellar pulsations. We first introduce the relevant equations and the practical steps required. We then review the major results to date, as well as emphasize important caveats in the use of this exoplanet discovery method.

\section{Pulsation Timing: Relevant Equations}

\runinhead{The practical concept} of using stellar pulsations to discover exoplanets arises from the dynamical perturbation of the oscillating star by an external companion, causing periodic light-travel-time changes detectable in the arrival times of otherwise stable pulsations. In the simplest case, an external companion on a circular orbit around a host star will cause light-travel-time delays with a semi-amplitude
\begin{equation}
A \simeq \frac{a \sin{i}}{c} \frac{m_p}{M_\star}  \;,
\label{eq:hermes1}
\end{equation}
where $a$ is the orbital semimajor axis with respect to inclination ($i$), $m_p$ is the mass of the companion, $M_\star$ is the mass of the pulsating star, and $c$ is the speed of light. Variation in light-travel time imparts a measurable phase change to the oscillations; the phase changes are often expressed as changes in the arrival times or the times of maximum brightness.

More generally, the timing changes are directly related to the radial velocity ($v_{\rm rad}$) of the host star. Following \citet{2014MNRAS.441.2515M}, the time delays are described by
\begin{equation}
\tau(t) = - \frac{1}{c} \int_0^t v_{\rm rad}(t') {\rm d}t' \;.
\end{equation}

It follows that the radial velocity of the host star can be described as
\begin{equation}
v_{\rm rad}(t) = - c \frac{{\rm d}\tau}{{\rm d}t} \;.
\label{eq:hermes3}
\end{equation}

Equation~\ref{eq:hermes3} can be numerically differentiated to directly derive the orbital properties through light-arrival-time delays \citep{2015MNRAS.450.4475M}. However, a more commonly used approach is to analytically describe the orbit as a Fourier series of sinusoidal functions:
\begin{equation}
\tau(t) = \sum_{k=1}^N A_k \sin{} (\frac{2 \pi k t}{P_{\rm orb}} + \phi_k) \;,
\end{equation}
where $A_k$ is the semi-amplitude, $P_{\rm orb}$ is the orbital period, and $\phi_k$ is the phase relative to a chosen zero point in time. This form directly translates to a periodogram analysis and can still accommodate eccentric orbits.

\runinhead{Orbital parameters} can thus be calculated. Pulsation timings can be directly converted into radial-velocity measurements, either to supplant existing data or to create an entire new set of observations.
The time-delay semi-amplitude ($A_1$) is directly related to the radial-velocity semi-amplitude ($K_1$), as it would be calculated from spectroscopy:
\begin{equation}
A_1 = \frac{2 \pi c}{P_{\rm orb}} K_1  \;.
\end{equation}

Importantly, the companion mass can be estimated if the mass of the host star can be accurately measured. The mass function is
\begin{equation}
f(M_\star, m_p, \sin{i}) = \frac{(m_p \sin i)^3}{(M_\star + m_p)^2} = \frac{(2 \pi)^2}{P_{\rm orb}^2 G} \, \frac{(a_1 \sin{i})^3}{(1-e^2)^{3/2}}  \;,
\end{equation}
where $G$ is the gravitational constant, $e$ is the eccentricity, and again $M_\star$ and $m_p$ are the masses of the host star and companion, respectively. This can be expressed as Equation~\ref{eq:hermes1} when the orbit is circular and $M_\star >> m_p$.

\runinhead{Phase delays are deduced in practice} using the ($O-C$) method, which compares the observed ($O$) pulsation time of maximum to the expected value calculated ($C$) assuming a constant pulsation period. In an ideal case, we have perfect knowledge of the underlying pulsation period, which is not changing in time, such that the only changes in the pulsation phase are due to a hypothetical external companion.

In reality, this is never the case. Fortunately, we can fit for all of these factors simultaneously. Adapted from the derivation of \citet{1991ApJ...378L..45K}, we can express the complete \omc\ equation as
\begin{equation}
O \; - \; C = t_{0} + \, \Delta P \, E + \, \frac{1}{2} P \dot{P} E^2 + \, A_1 \sin \left(\frac{2 \pi E}{P_{\rm orb,1}} + \phi_1 \right)
\label{eq:hermes8}
\end{equation}
where $t_0$ is the reference phase, $\Delta P$ is the overall uncertainty in the pulsation period, $\dot{P}$ corresponds to any secular pulsation period change, $\phi_1$ is the phase-variation phase, and $A_1$ is the amplitude of the phase variations caused by the external companion. All times are referenced as an integer number of epochs ($E=t/P$), which is simply the observing time ($t$) divided by the pulsation period ($P$).

\section{Pulsation Timing: Practical Methodology}

\runinhead{Choosing the appropriate class of variable star} for pulsation timing studies is the first observational step. Stars for which the pulsation timing method yields the highest signal-to-noise are those with the highest-amplitude, shortest-period oscillations. Equally important, the pulsations must have exceptional intrinsic phase stability; more precisely, the mode lifetimes must greatly exceed the orbital period of the companion. Stochastically driven pulsations in solar-like oscillators are thus poor clocks for timing studies, which exclude most cool dwarfs and giants from this technique.

\begin{figure}
\includegraphics[width=1.0\textwidth]{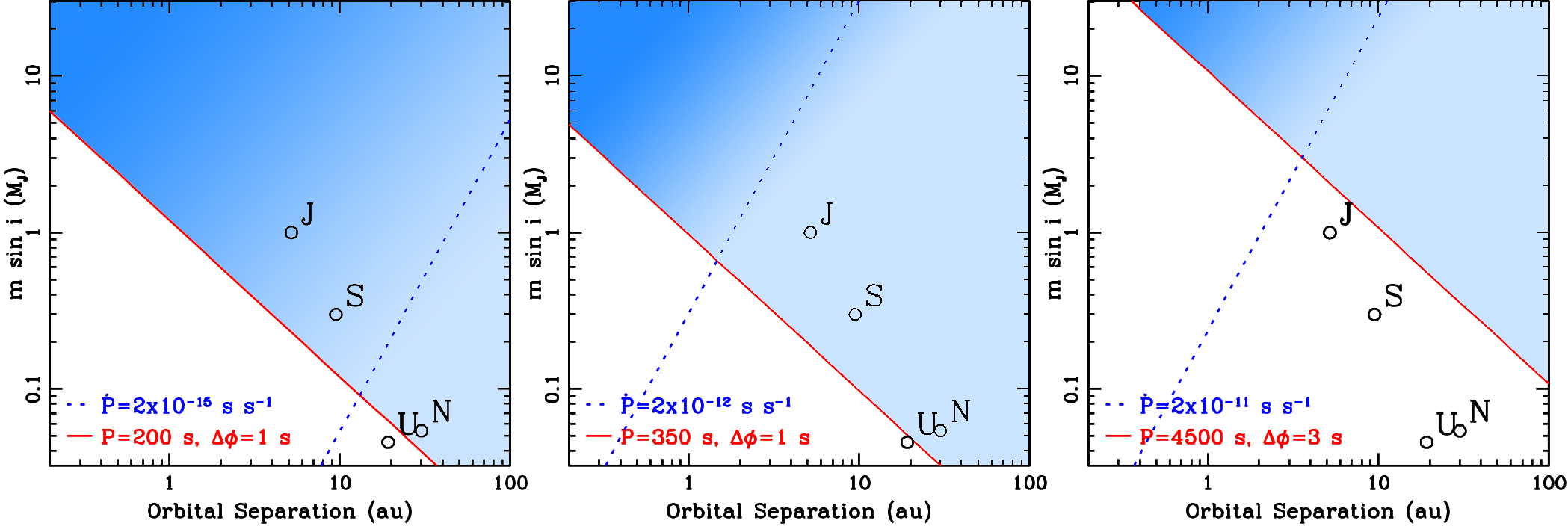}
\caption{Shaded areas show the region of detectability (as a function of companion mass vs. semimajor axis) for companions in circular orbits around a typical 0.6 \msun\ white dwarf (left), a 0.47 \msun\ hot subdwarf (middle), and a 1.8 \msun\ A star (right). Giant planets in our solar system are labeled. The short-period modes of white dwarfs and subdwarfs provide the most sensitivity. Additionally, typical rates of period change ($\dot{P}$) from stellar evolution are shown, which must be accounted for to detect the longest-period systems.}
\label{fig:hermes1}      
\end{figure}

Motivated by the thousands of pulsating stars with 4 years of high-precision photometry from the {\em Kepler} space telescope, \citet{2016MNRAS.461.1943C} explored the objects with the greatest sensitivity to low-mass companions. They found that $\delta$ Scuti variables (A stars exhibiting pressure-mode pulsations), pulsating hot subdwarfs (sdBs), and pulsating white dwarfs provided the greatest sensitivity. Using typical values of pulsation periods, stellar mass, and phase uncertainties for an epoch of observations, we show in Figure~\ref{fig:hermes1} general sensitivity limits for pulsation timing of these three types of stars. The figures emphasize how the sensitivity increases dramatically the longer a star is monitored.

\runinhead{Measurements of phase delays} require a long baseline of observations, covering at least one orbit of the host star around the center of mass of the system. In practice, this is accomplished by cultivating multiple seasons of time-series photometry of the pulsating star (or breaking up near-continuous photometry from space-based missions such as {\em Kepler} or {\em TESS} into shorter subsets). Importantly, since the amplitude of the light-travel times caused by exoplanets is of order a few seconds or below, the time stamps for each exposure must be corrected to the barycenter of our own solar system; the history and practical steps for this process are well described by \citet{2010PASP..122..935E}.

Using the whole dataset, the periods of stellar variability are deduced to high precision, usually using a nonlinear least-squares fit for the frequency, amplitude and phase of the pulsations present in the star. Subsequently, a simultaneous linear least-squares fit (holding the frequency fixed) is performed on each subset, producing a phase measurement at each epoch of observation. All modes must be fitted simultaneously within each epoch in order to account for interaction (such as beating) between pulsation modes closely spaced in frequency. These phase measurements (usually the time of pulsation maximum) observed as a function of time construct an \omc\ diagram for each pulsation mode present in the star.


A periodogram of the \omc\ diagram for each pulsation mode can then be used to search for binary modulation, after fitting out any linear trends that arise from an imprecise pulsation period or any parabolic trends caused by secular period change in the oscillation (see Equation~\ref{eq:hermes8}). A significant peak in this periodogram can indicate the presence of an external companion; the frequency is directly commensurate with the orbital frequency, and the amplitude of the peak relates to the semi-amplitude of the light-travel-time variation. The amplitude of the primary peak relative to its harmonics can also be used to constrain the eccentricity of the orbit (\citealt{2014MNRAS.441.2515M}, following \citealt{2012MNRAS.422..738S}):
\begin{equation}
e \simeq \frac{2A_2}{A_1} \simeq \frac{4A_3}{3A_2} \;,
\end{equation}
where $A_1$ is the amplitude of the primary peak in the periodogram and $A_2$ and $A_3$ are the amplitudes of the first and second harmonics, respectively.

Light-travel-time effects must affect all modes identically, establishing the case to inspect the \omc\ diagrams of all modes in the star to distinguish an internal effect in the star from one caused by an external companion.

\section{Pulsation Timing: Observational Results}

\runinhead{Stellar binary companions} have been the most common type of previously unseen objects discovered through pulsation timings; more massive objects cause a larger change in the pulsation arrival times, making them easier to detect.

A robust verification of the pulsation timing method was observed in fortnightly fluctuations in the \omc\ diagram of photometry from the hot subdwarf CS 1246 \citep{2011MNRAS.414.3434B}, which were confirmed spectroscopically \citep{2011ApJ...737L...2B}. Both the $\tau = 10.7\pm0.4$ s pulsation-mode phase variations and the $16.6\pm0.6$ \kms\ radial-velocity variations observed every 14.103 days are independently consistent with an $m_2 \sin{i}=0.129\pm0.005$ \msun\ companion, likely a low-mass M dwarf outshone by the $>$$28{,}000$-K pulsating sdB.

With a long baseline of nearly continuous observations, many other binaries have been discovered around pulsating stars observed during the original {\em Kepler} mission. Hundreds have been discovered around the numerous $\delta$ Scuti stars \citep{2013MNRAS.432.2284M,2014MNRAS.443.1946B,2016MNRAS.461.1943C}, including many of which that have been verified from radial-velocity follow-up \citep{2016MNRAS.461.4215M}. 
In addition, multiple hot subdwarfs have revealed stellar companions from timing their pulsations \citep{2012AnA...544A...1T,2014AnA...570A.129T}.

\runinhead{Only a handful of exoplanets} have been claimed using the pulsation timing method.

\begin{figure}
\includegraphics[width=1.0\textwidth]{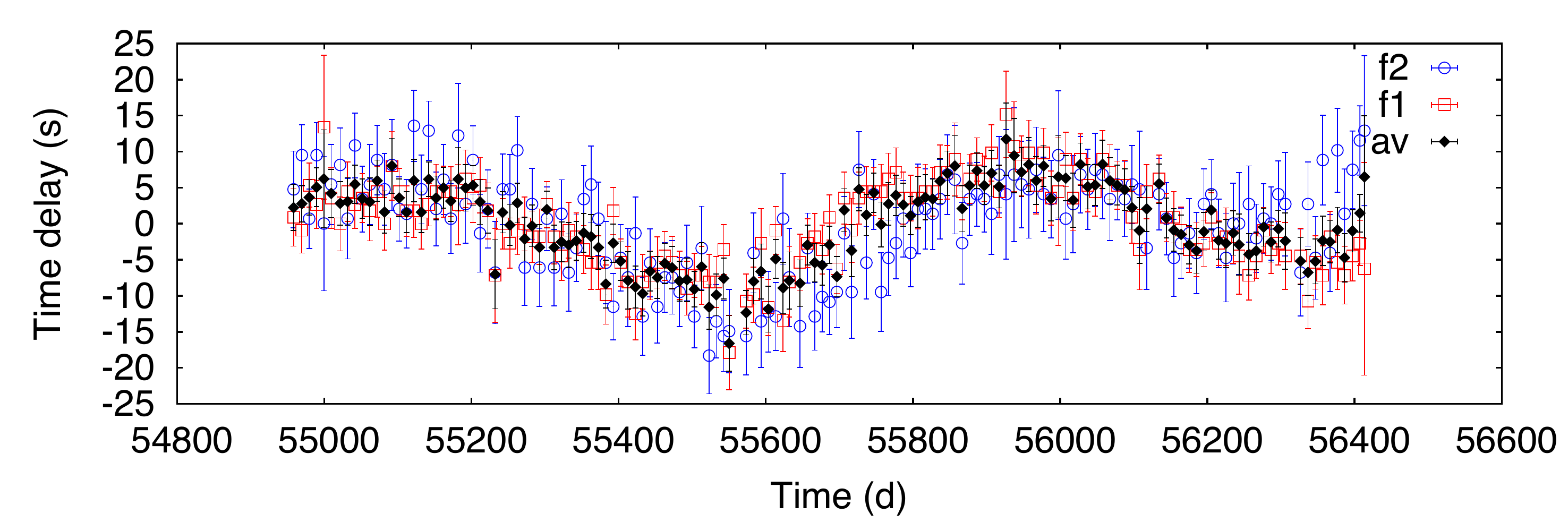}
\caption{Time delays in the two largest-amplitude pulsations of the $\delta$ Scuti star KIC 7917485 are consistent with an $m \sin{i} = 11.8^{+0.8}_{-0.6}$ M$_{\rm Jup}$ object in an $840^{+22}_{-20}$ day, slightly eccentric orbit. Reproduced from \citet{2016ApJ...827L..17M} by permission of the AAS.}
\label{fig:hermes2}      
\end{figure}

Most secure is the object likely near the deuterium-burning limit that orbits near the habitable zone of an A star in the original {\em Kepler} mission: an $m \sin{i} = 11.8^{+0.8}_{-0.6}$ M$_{\rm Jup}$ object in an $840^{+22}_{-20}$ day, slightly eccentric ($e=0.15^{+0.13}_{-0.10}$) orbit around the $\delta$ Scuti star KIC 7917485 \citep{2016ApJ...827L..17M}. Figure~\ref{fig:hermes2} shows time delays of the two highest-amplitude pulsation modes, which have periods of 1.56 hr and 1.18 hr, along with their weighted averages. Both pulsations show an identical phase modulation with an amplitude of $\tau=7.1\pm0.5$ s that reveals a companion close to the brown-dwarf-planet boundary. Interestingly, the planet has an insolation flux between roughly $1.2-1.4$ times that of the Earth, such that any moons around this object could potentially be habitable \citep{2016ApJ...827L..17M}.

The least-massive object claimed using the pulsation timing method was proposed by \citet{2007Natur.449..189S} around the hot subdwarf V391 Pegasi. This sdB shows evidence of a $\tau=5.3\pm0.6$ s phase modulation every $1170\pm44$ days in two pulsations (with periods of 349.5 s and 354.1 s), which would correspond to the effect of an $m \sin{i} = 3.2 \pm 0.7$ M$_{\rm Jup}$ planet. Additional substellar companions have been proposed to orbit two other hot subdwarfs with long-term monitoring \citep{2012AN....333.1099L}.

\runinhead{A major caveat} in the use of pulsation timing to detect exoplanets is the inherent stability of the pulsation modes themselves. Here the white dwarfs, some of the most sensitive objects for detecting substellar companions using pulsations, serve as a useful lesson.

In the early 2000s, researchers at McDonald Observatory undertook a program to monitor more than a dozen of the hottest pulsating white dwarfs with hydrogen atmospheres, since these stars have the longest mode lifetimes and thus the most stable pulsations \citep{2003ASPC..294...59W}. Early on, one white dwarf in this sample, GD 66, showed phase modulation that could be caused by a $\sim$2 M$_{\rm Jup}$ planet in a 4.5-yr orbit \citep{2008ApJ...676..573M}. However, this modulation came from just one pulsation mode. Subsequent observations have shown that different modes in the same star show different phase modulation; this excludes an external cause for the changes, complicating the planetary hypothesis for GD 66 \citep{2013AAS...22142404H,2013PhDT.......170D}.

Critically, multiple white dwarfs with otherwise ``stable'' oscillations have shown unexpected changes in the pulsation arrival times: one shows phase evolution hundreds of times faster than expected from stellar evolution \citep{2013ApJ...766...42H}, another shows divergent phase modulation in multiple different pulsation modes that cannot be caused by an external companion \citep{2013ApJ...765....5D}, and yet another shows phase modulation within the same rotational multiplet that is anticorrelated, which again cannot be caused by an external companion \citep{2016AnA...585A..22Z}.

The authors of the latter result suggest nonlinear resonant mode coupling could explain the phase changes. Regardless of the physical cause, these three independent empirical results warn that internal effects must be ruled out as the source of the phase modulation when using the pulsation timing method to discover exoplanets. This is best accomplished by observing the phase evolution of as many modes as possible, to guarantee they all respond to a hypothetical external companion in the same way.

\section{Conclusions}

Exoplanets may be revealed around pulsating stars by carefully monitoring the arrival times of the pulsations, searching for periodic modulation from the light-travel-time delays caused when the host star is tugged by an unseen companion. The method requires the pulsations be extremely stable, and is most sensitive to substellar companions around A stars and compact remnants. Additionally, phase modulation should be observed in more than one mode within the star to exclude effects caused by changes to the stellar interior --- external effects from an exoplanet or binary companion identically affect the arrival times of all pulsation modes.
Hundreds of binary companions have been revealed using the pulsation timing method, which impart larger signals. But exoplanets are not beyond reach --- several substellar companions have been discovered by monitoring the arrival times of stellar oscillations.



\begin{acknowledgement}
The author thanks Simon Murphy and Bart Dunlap for helpful discussions. Support for this work was provided by NASA through Hubble Fellowship grant \#HST-HF2-51357.001-A.
\end{acknowledgement}


\end{document}